\documentclass[byrevtex,pre,preprint,showpacs]{revtex4}
\usepackage{graphicx}
\usepackage{amsmath}
\newcommand{\fig}[1]{Fig.~\ref{#1}}
\newcommand{\eq}[1]{Eq.~(\ref{#1})}
\begin{document}
\title{Transport properties of incipient gels}

\author{Sune N\o rh\o j Jespersen}
\email[]{sjespers@sfu.ca}
\affiliation{Department of physics, Simon Fraser University, Burnaby,
British Columbia, Canada V5A 1S6}
\author{Michael Plischke}
\email[]{plischke@sfu.ca}
\affiliation{Department of physics, Simon Fraser University, Burnaby,
British Columbia, Canada V5A 1S6}

\date{\today}

\begin{abstract}
We investigate the behavior of the shear viscosity $\eta(p)$ and the
mass-dependent diffusion coefficient $D(m,p)$ in the context of a simple
model that, as the crosslink density $p$ is increased, undergoes a
continuous transition from a fluid to a gel. The shear viscosity
diverges at the gel point according to $\eta(p)\sim (p_c-p)^{-s}$ with
$s\approx 0.65$. The diffusion constant shows a remarkable dependence
on the mass of the clusters: $D(m,p)\sim m^{-0.69}$, not only at $p_c$
but well into the liquid phase. We also find that the Stokes-Einstein
relation $D\eta\propto k_BT$ breaks down already quite far from the
gel point.
\end{abstract}

\maketitle

\section{\label{sec.intro}Introduction}
When a system of polyfunctional molecules is crosslinked, the transport
properties such as the shear viscosity and the diffusivity can be
dramatically affected. In particular,  the diffusivity decreases as the
number of crosslinks is increased and the shear modulus increases, diverging
at the critical point at which a gel is formed. The diffusivity remains
finite as the system gels since monomers and small clusters can diffuse
through the tenuous structure that characterizes the amorphous solid close to
the critical point. Although gels have been studied for many years
\cite{adam96}, their critical behavior remains poorly understood. In
particular the question of whether or not there exist universality classes
into which different materials can be grouped remains largely unanswered.\par
In this article, we report on extensive molecular dynamics simulations of a
simple model for a gel. We study the system on the fluid side of the gel
point from the simple liquid limit into the critical region. We investigate
the structural properties of clusters and calculate both the shear viscosity
$\eta(p)$ and the mass-dependent diffusion constant $D(m,p)$ as function of
the crosslink density $p$. We find that as $p\to p_c$, $\eta(p)\sim
(p_c-p)^{-s}$ with $s\approx 0.65$, a value somewhat smaller than that
conjectured by de Gennes \cite{deGennes79} on the basis of an analogy with
a random superconductor network and also predicted recently by Broderix {\it
et. al.} \cite{broderix99} for a Rouse-like model network. The mass-dependent
diffusion constant $D(m,p)\sim m^{-0.69}$ for a range of $p$ near the
critical point and $3\leq m\leq 50$. This behavior is consistent with earlier
results for $p=p_c$ \cite{sj02} and rather close to a prediction
\cite{deGennes79} made on the basis of a simple scaling argument. On the other
hand, our value for $s$ is somewhat lower than the one found by K\"untzel
\textit{ et. al.} in a recent article \cite{kuntzel03} in which $s=0.8$ is
found by a theoretical analysis of the Zimm model. The
 diffusion coefficient $D(m,p)\to  {\rm const.}$ as $p\to p_c$ for  $m$  at
least as large as 10 but displays critical behavior in the next leading term.
It is also worth noting that, in contrast to simple liquids, the product
$D(p)\eta(p)$ is not a constant but rather reflects the divergence of $\eta$
at the gel point.\par
The structure of this article is as follows. In section
\ref{sec.model} we describe our model and the computational details. Section
\ref{sec.static} contains a discussion of the geometric properties of the
clusters and the nature of the percolation transition. The shear viscosity
calculation and results are described in section \ref{sec.visc} and results
for the diffusion constants are found in \ref{sec.dif}. We conclude with a
short discussion in section \ref{sec.con}.


\section{\label{sec.model} Model}
The model is similar to the one employed in \cite{sj02}, but we
include the details below for completeness.
Our system is composed of $N=L^3$ ($L=10,13,15,20$ and $30$) particles
interacting pairwise through the shifted
Lennard-Jones potential
\begin{equation}
U({\mathbf r})=\begin{cases}
U_{LJ}(r)-U_{LJ}(2.5\sigma), & r \leq 2.5\sigma\\
0 & \text{otherwise,}\end{cases}
\end{equation}
where
$U_{LJ}(r)=4\epsilon((\sigma/r)^{12}-(\sigma/r)^{6})$. All of our
simulations are $3D$ constant energy molecular dynamics (MD) simulations
corresponding to an average temperature of $k_BT/\epsilon\approx 1$ and
density
$\Phi=0.8\sigma^{-3}$. These
choices ensure that the system is in the liquid-phase region of the
phase-diagram \cite{smit92,felicity97}. We use periodic boundary conditions
and a
time step of magnitude $dt=0.005\tau$, where
$\tau=\sqrt{m\sigma^2/\epsilon}$ is the reduced Lennard-Jones time. From a
typical equilibrium state of this liquid we let the particles form a
specified number $n$  of permanent chemical bonds if they come closer than
$r_c=2^{1/6}\sigma\approx 1.12$, coinciding with the minimum of $U(r)$.
The bond interaction is a harmonic oscillator
potential
$U_{\text{harm}}(r)=1/2 kr^2$: in our simulations we take
$k\sigma^2/\epsilon=2.0$ (different from \cite{sj02}). Note that this way
of adding bonds violates energy
conservation; indeed we actually pump energy into the system when
adding bonds. To compensate we cool down the system again after having
established the required number of bonds.
With
this bonding procedure cross linking is very fast --- the average distance
between the particles is comparable  to $r_c$, so a large number of
particles are available for bonding at any given instant.
Each particle can bond to a maximum of $f=6$ other 
particles (excluding itself), and the cross link density $p$ is then given in
terms of the number of bonds $n$ as $p=2n/fN$. Any number of particles, if
fulfilling 
the conditions above, can be cross linked per time step, but we halt
the bond formation when $p$ reaches a predetermined value.

\section{\label{sec.static}Geometric properties}
\begin{figure}
\includegraphics{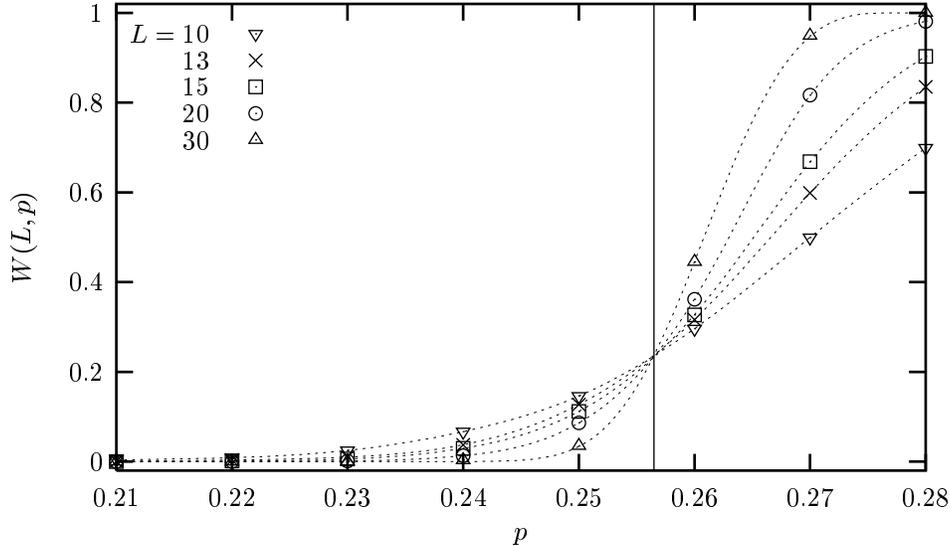}
\caption{\label{fig.W} Fraction of systems $W(L,p)$ percolating in the
$x$-direction as a function of $p$ and for five different system sizes
as indicated on the plot. The lines are guides for the eye,
except in the case $L=30$  for which the data are fitted to a stretched
 exponential \cite{newman01}. We estimate $p_c=0.2565$.}
\end{figure}

\begin{figure}
\includegraphics{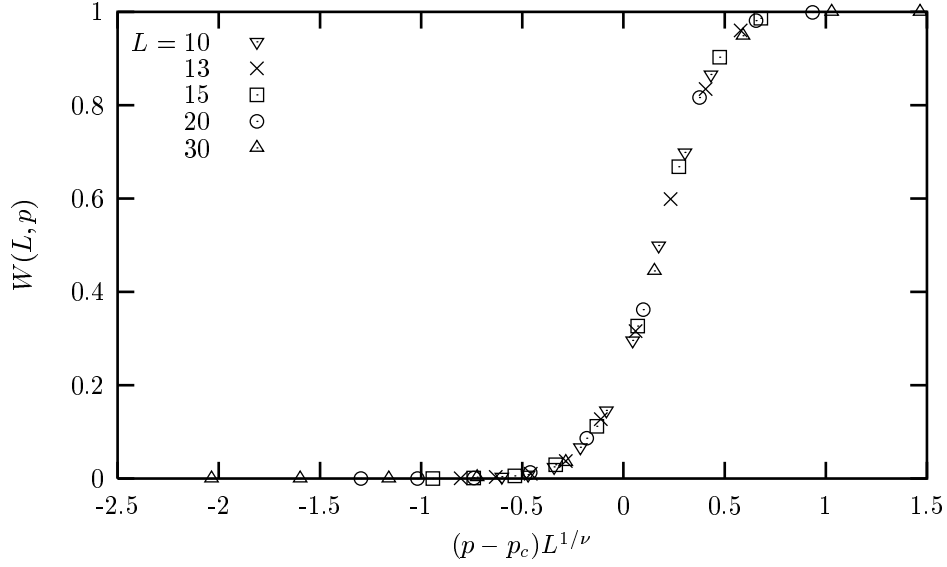}
\caption{\label{fig.Wscaling} Same as \fig{fig.W}, except here
$W(L,p)$ is plotted as a function of $L^{1/\nu}(p-p_c)$ with
$p_c=0.2565$ and $\nu=0.9$. The data collapse very nicely in agreement
with finite size scaling theory.}
\end{figure}

Before discussing the dynamic properties of this model, we need some basic
information about the static properties. In this section we determine the
geometrical percolation point $p_c$ as well as the two critical
exponents $\nu$, the correlation length exponent, and $\gamma$ the
exponent characterising the divergence of the weight average cluster mass
(of finite clusters).
We follow a similar procedure to the one used and outlined in
\cite{sj02,plischke02,vernon01}. In
order to find $p_c$ we calculate numerically the fraction $W(L,p)$ of
percolating systems of size $L^3$ with a bond density $p$. This
function is plotted in \fig{fig.W} for all five system sizes. The
crossing points of the different curves seem to coincide, and the
corresponding value of $p$ is thus a good estimate of $p_c$ \cite{sj02,ziff02}:
 From the
figure we determine $p_c=0.2565$ as in \cite{sj02}.
Finite size scaling theory predicts that $W(L,p)$ does not depend on
$L$ and $p$ separately but only on the combination $L/\xi$ (and the
sign of $p-p_c$) where
$\xi=|p-p_c|^{-\nu}$ is the correlation length
and $\nu$ the correlation length exponent \cite{stauffer92}. Thus we
may write
\begin{equation}
\label{sizescale}
W(L,p)=f(L^{1/\nu}(p-p_c)),
\end{equation}
where $f(x)$ is a scaling function.To test this hypothesis we
replot the data for $W(L,p)$ from \fig{fig.W} in \fig{fig.Wscaling} as
a function of $L^{1/\nu}(p-p_c)$ with $p_c=0.2565$ as determined above and
$\nu=0.9$.
The collapse is very good confirming the correctness of the values for
$p_c$ and $\nu$.

\begin{figure}
\includegraphics{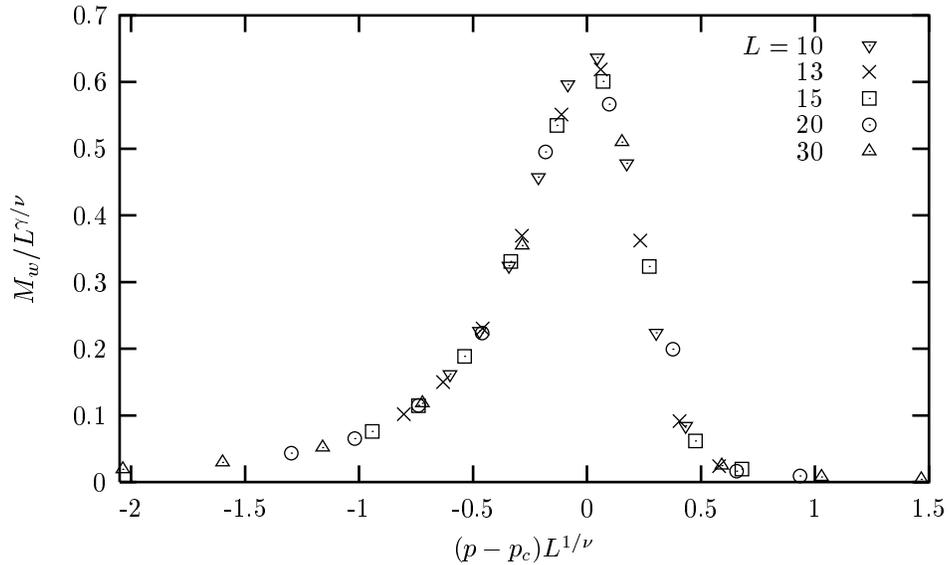}
\caption{\label{fig.mwscal} Scaling plot of the weight average molecular
weight $M_w$. The quality of the data collapse confirms $\gamma=1.8$
in accordance with the $3D$ percolation value.}
\end{figure}

To compare with percolation theory we need one more exponent, and here
we consider the behavior of the weight average cluster mass $M_w$. In
the thermodynamic limit the expected behavior is $M_w(p)\sim
|p-p_c|^{-\gamma}$ \cite{stauffer92}: Therefore we compute $M_w$ as a
function of $p$ for different system sizes, and in \fig{fig.mwscal} we plot
the results in the form $M_w/L^{\gamma/\nu}$ versus
$L^{1/\nu}(p_c-p)$ with $\gamma=1.8$ being the expected $3D$
percolation value and $\nu$ and $p_c$ as determined previously. Again
there is a very nice data collapse, and we therefore conclude that
here as in \cite{sj02,plischke02,vernon01}  our system is consistent
with the $3D$  percolation universality class in so far as static
properties are concerned.

\section{\label{sec.visc}Viscosity}
We measure the shear viscosity $\eta(p)$ by using the appropriate
Green-Kubo formula \cite{Allen, Hansen}:
\begin{equation}
\label{eta}
\eta=\frac{1}{V k_B T}\int_0^\infty \, dt \langle
\sigma_{xy}(t)\sigma_{xy}(0)\rangle,
\end{equation}
where $V$ is the volume and $\sigma_{xy}(t)$ the $xy$ component of the
stress tensor:
\begin{equation}
\sigma_{xy}(t)=\sum_{i=1}^N m v_{x,i}v_{y,i}+
\sum_{i=1}^N \sum_{j<i}(y_i-y_j)f_{x,ij}.
\end{equation}
In this equation $f_{x,ij}$ is the $x$ component of the force from
particle $j$ on particle $i$, and the meaning of the remaining terms
is self explanatory. In the simulations we average over several
hundred samples for each $p$ and over three
off-diagonal components ($xy,yz$ and $zx$) of the stress tensor to obtain
slightly better
statistics. It is important to note that we have discarded any sample
containing a spanning cluster since for such a system the viscosity is not
defined, i.e.\ the right hand side of \eq{eta} diverges.  Although we have
simulated very long runs (up to
$t=750\tau$) the stress correlator $ C_{\sigma\sigma}(t)\equiv \langle
\sigma_{xy}(t)\sigma_{xy}(0)\rangle$ has still not decayed completely,
and it is necessary to add by hand an additional contribution, in
particular for $p$ close to $p_c$. A stretched exponential
$C_{\sigma\sigma}(t)= a \exp(-bt^c)$ with $0.1<c<0.3$ seems to fit the data
well for
long times, and there are also theoretical reasons \cite{broderix01} to
believe that this is
the appropriate form. See \cite{vernon01} for a thorough discussion of this
point.

\begin{figure}
\includegraphics{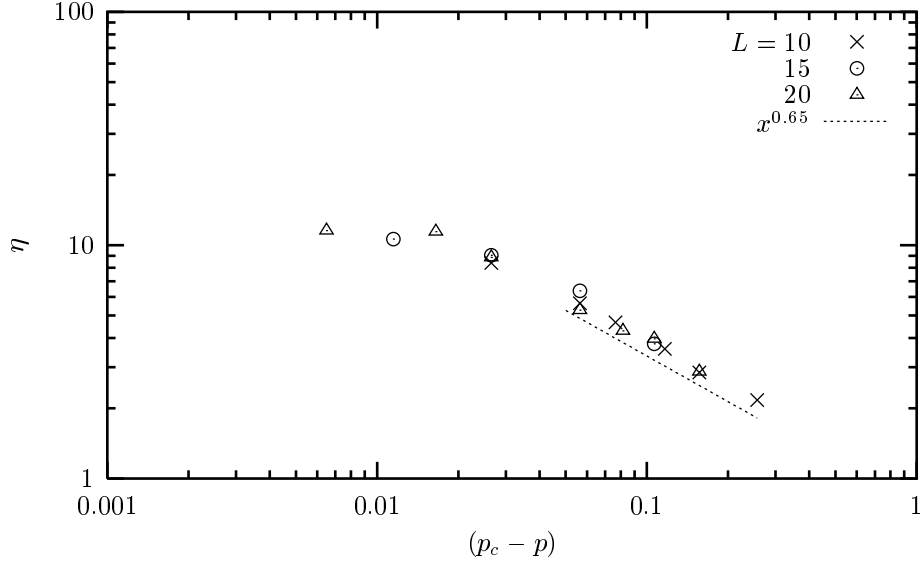}
\caption{\label{fig.visc} The dimensionless shear viscosity as a function of
$p_c-p$ for
  different $L$.}
\end{figure}
\begin{figure}
\includegraphics{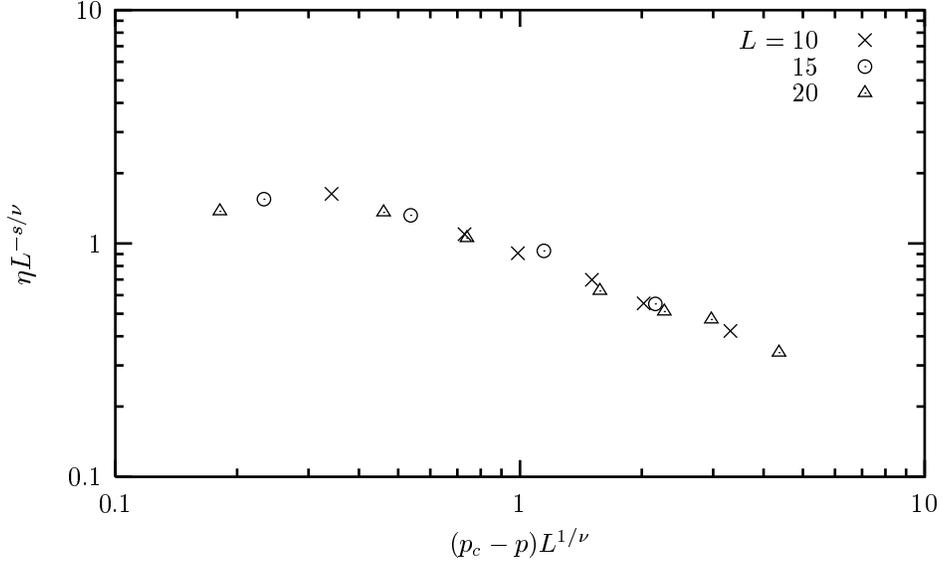}
\caption{\label{fig.visc_scal} Same as \fig{fig.visc}, but here plotted in a scaling form with $s=0.65$.}
\end{figure}

In \fig{fig.visc} we have plotted the resulting values for the
viscosity for different systems sizes and at different stages of the
cross linking. We note the clear power law behavior outside the critical
region, and a fit to the $L=10$ data in this region  yields $s=0.65$. The line $\eta\propto
(p_c-p)^{-0.65}$ has also been drawn on the plot, and it is apparent that the
data are consistent with this exponent. For large $p$, $p>0.23$, there are
larger error bars and this will also affect the scaling plot.
Since the viscosity diverges at the critical
point with an exponent $s>0$, the finite size scaling form is
\begin{equation}
\eta(p,L)=L^{-s/\nu}g(\xi/L) \quad p<p_c,
\end{equation}
where $g$ is a scaling function with the limits
\begin{equation}
  g(x)\propto
  \begin{cases}
    x^{-s/\nu} & x\to 0\\
    \text{const.} & x\to \infty,
  \end{cases}
\end{equation}
and $\xi\sim(p_c-p)^{-\nu}$ is the
correlation length, c.f. Sec.~\ref{sec.static}. Therefore we plot in
\fig{fig.visc_scal} $\eta L^{s/\nu}$ versus $L^{1/\nu}(p_c-p)$, and the
collapse is quite good outside the critical region with $s=0.65$, whereas
there is a larger scattering of the points for $p$ closer to $p_c$.

de Gennes has suggested \cite{deGennes79} a value of $s\approx 0.7$ based
on an analogy
between gelation and conductance in a random mixture of normal and
superconducting elements, and nice agreement with this was found in a
related model in \cite{vernon01}. Here we have observed a slightly smaller
value for $s$.

\section{\label{sec.dif}Diffusion}
In this section we extend our earlier study \cite{sj02} on the diffusion
of clusters. Previously we were concerned mainly with the behaviour of
the diffusion constant $D(m,p)$ as a function of cluster mass $m$ at
the gelation point $p=p_c$. Here we address the $p$
dependence of $D$ for different clusters and the validity of the
Stokes-Einstein relation  $D(p)\propto k_BT/\eta(p)$ for a
given cluster mass. We restrict our attention to the $L=20$ system.

To determine the diffusion constant we use the Einstein relation:
\begin{equation}
\label{einstein}
  \frac{1}{6t} \langle (r_m(t)-r_m(0))^2 \rangle \xrightarrow{t\rightarrow
\infty} D(m,p),
\end{equation}
where $r_m(t)$ is the center-of-mass position of a cluster of mass $m$
at time $t$, for a given value of $p$ (for clarity of the presentation we omit
the explicit dependence on $p$ in the notation). When calculating the
diffusion constant numerically we have averaged over all clusters of a
given mass $m$ and over several hundred crosslinkings, and we have
discarded any percolating samples. This has been done mainly for
consistency when comparing with $\eta$, but in any event we do not
expect this to affect the diffusion of any but the very largest clusters.

\begin{figure}
\includegraphics{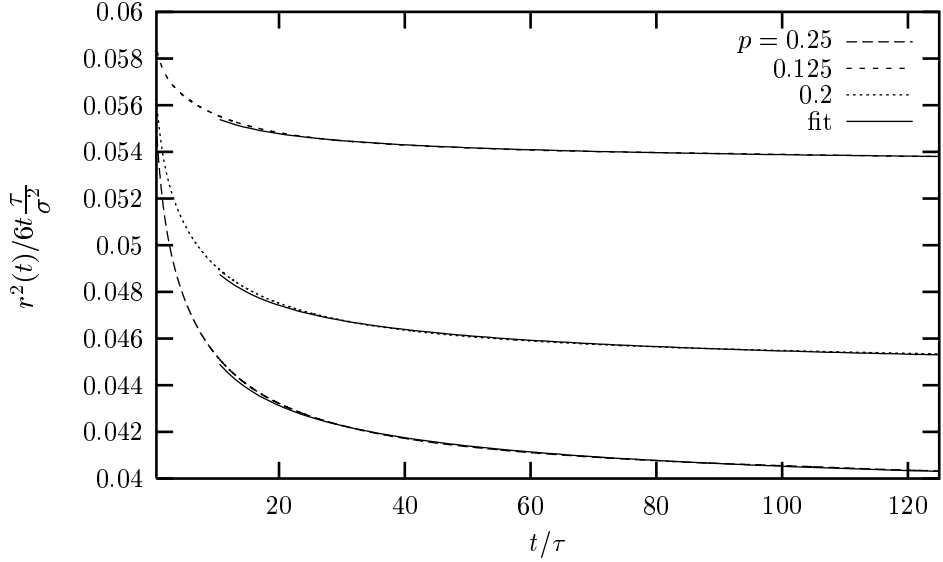}
\caption{\label{fig.tail1} $\langle (r_1(t)-r_1(0))^2\rangle /6t$ a
function of time for monomers,  and for  three different values
of $p$:
    $0.125$, $0.2$  and $0.25$ from top to bottom.  The long time
    tails are clearly visible, and the solid lines are fits to
    the same functional form (see text).}
\end{figure}

\begin{figure}
\includegraphics{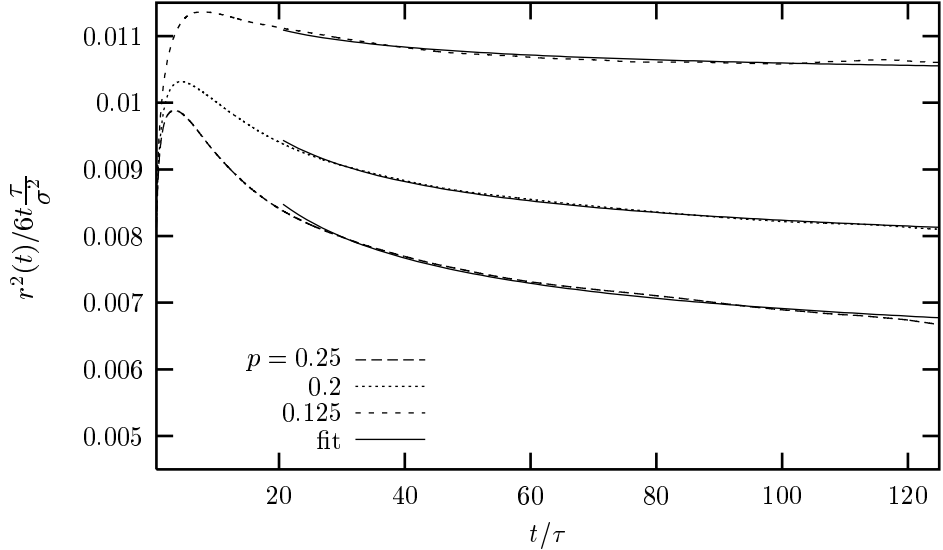}
\caption{\label{fig.tail2} Same as \fig{fig.tail1} but this time for
    clusters of size $10$.}
\end{figure}

First we examine the convergence of \eq{einstein} by plotting in
\fig{fig.tail1} the behavior of $\langle (r_m(t)-r_m(0))^2\rangle /6t$
for $m=1$ (monomers) as a function of time and for three different values of
$p$.
From these curves we clearly see the existence of long-time tails in the
velocity auto-correlation function. Consider the ``Green-Kubo'' formula
corresponding to \eq{einstein}:
\begin{equation}
\label{difgk}
  \frac{\langle (r_m(t)-r_m(0))^2\rangle}{t}= \int_0^t ds\,
  \langle \mathbf{v}_m(s)\cdot \mathbf{v}_m(0) \rangle (1-s/t).
\end{equation}
The dominant contribution to  $\langle (r_m(t)-r_m(0))^2 \rangle /t$
at large times is
\begin{equation}
\label{difasymp}
 \frac{\langle (r_m(t)-r_m(0))^2\rangle}{t} = D(m,p)-\int_t^\infty ds\,
  C^{(m)}_{vv}(s).
\end{equation}
where $C^{(m)}_{vv}(s)=\langle \mathbf{v}_m(s)\cdot \mathbf{v}_m(0) \rangle$ is
the velocity auto-correlator. Therefore, a power law tail
$C^{(m)}_{vv}(s)\sim t^{-\alpha}$ in the velocity auto-correlation function
will translate into a corresponding power law tail $\langle
(r_m(t)-r_m(0))^2 \rangle /t \sim D(m,p)+\text{const.}\,t^{1-\alpha}$ in the
Einstein relation. In simple liquids a value of $\alpha=3/2$ is
ubiquitous \cite{Hansen}, and has also been observed for gelating systems in
\cite{vernon01}: here we find that the same power-law provides a very good
fit to the data for all $m$, but in particular for the small
clusters. In  \fig{fig.tail1} we have also plotted these fits to the
power-law $a+bt^{-1/2}$, and the deviation from the simulation results
at early times
is barely visible. In figure \fig{fig.tail2} we have done the same for
clusters of mass $10$, and we see the same behavior. The agreement is
slightly worse, presumably due to poorer statistics of larger
clusters.
We note the existence of a maximum in all of the curves (though not
visible on \fig{fig.tail1} for $m=1$) for $\langle (r_m(t)-r_m(0))^2 \rangle
/t$. By differentiating \eq{difgk} this can be shown to occur at $t_m$, where
$t_m$ is the solution to
\begin{equation}
 \int_0^{t_m}ds\, C^{(m)}_{vv}(s)s =0.
\end{equation}
An obvious consequence of the fact that for $t>t_m$ (using \eq{difasymp})
\begin{equation}
 \frac{d}{dt}\frac{\langle (r_m(t)-r_m(0))^2\rangle}{t} \approx
C^{(m)}_{vv}(t) < 0
\end{equation}
 is that $C^{(m)}_{vv}(t)$ becomes negative
(anti-correlation) for large $t$ and stays negative thereafter. This means that
the $1/\sqrt{t}$ tails in Figs.~\ref{fig.tail1}--\ref{fig.tail2} correspond to
a {\em   negative} $t^{-3/2}$ tail in $C^{(m)}_{vv}(t)$. We also note that
$t_m$ is an increasing function of $m$ and a decreasing function of $p$.

The error made by taking $D(m,p)$ to be the value of
$\langle(r_m(t)-r_m(0))^2 \rangle /6t $ at the end of the simulation
time $t=120\tau$ is neglibly small: for $p=0.2$ we have compared with
simulations that are twice as long, and at least for the $20$ lightest clusters
for which we had good enough statistics, the error was less than
$5\%$. For the smallest $m$ where the statistics are very good, one
can also obtain $D(m,p)$ from a power-law fit as mentioned above, and
the outcome is still consistent with the statement just made (the
error here is even much smaller than $5\%$).

In \cite{sj02} we studied $D(m,p_c)$ and we found the power-law
$D(m,p_c)\sim m^{-0.69}$. We have repeated this study up to clusters
of size $50$, and we observe the same behavior over the entire
range. For $p< p_c$, we see the same power law as a function of
$m$, at least for small cluster sizes. The quality of the statistics for larger
cluster sizes is insufficient to determine whether there is a
cross-over or cut-off as $m\to m^*(p)$, where $m^*(p)\sim
(p_c-p)^{-1/\sigma}$ is the mass of the largest cluster, but it seems
likely that there is. de Gennes has argued that for masses $1<m<m^*(p)$,
$D(m)\sim m^{-(\nu+s)/(\beta+\gamma)}$ on the basis of a Stokes-Einstein
relation with a mass dependent viscosity \cite{deGennes79}. Here $\beta$ is
the exponent that describes the decrease of the order parameter near
percolation:
$x_{\text{gel}}\sim (p-p_c)^{\beta}$, where $x_{\text{gel}}$ the
fraction of particles on the spanning cluster and $p\to p_c+$. The other
exponents have been
introduced already. By using the appropriate scaling relations for $3D$
percolation
\cite{stauffer92}, the exponent can be rewritten so the prediction is
$D(m)\sim m^{-2(\nu+s)/(d\nu+\gamma)}$, where $d=3$ is the Euclidian
dimension. With our values for the remaining exponents we get:
\begin{equation}
  D(m,p)\sim m^{-0.69}\quad\text{for}\quad 1<m<m^*(p).
\end{equation}
This is in very good agreement with our simulation results within
the observed power law regime. The theoretical prediction can be
rewritten as $D(R_g)\sim R_g^{-(1+s/\nu)}$ where $R_g\sim m^{1/D_f}$ is
the radius of gyration and $D_f$ the fractal dimension. This form of the
relation has sometimes (see for example \cite{gado00}) been used to infer
$s$ from the scaling of $D$
with $R_g$, but to the best of our knowledge the present study
presents the first direct verification of such a link.

\begin{figure}
\includegraphics{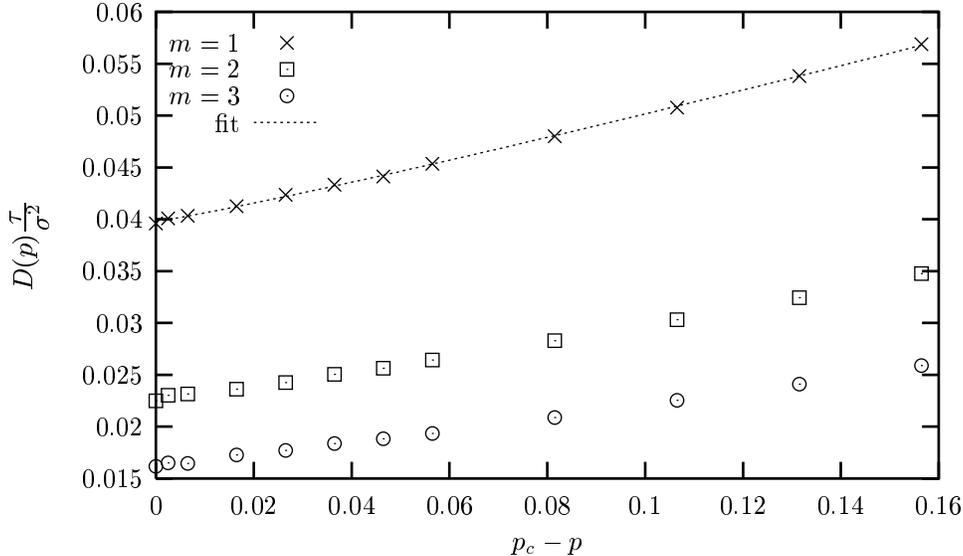}
\caption{\label{fig.Dvsp} Diffusion constant as a function of $p$ for
  clusters of three different sizes: $m=1$, $m=2$ and $m=3$ from top
  to bottom. The solid line is a fit to the function
  $D(p)=a(p_c-p)^{b}+D_c$ and $D_c=0.0398$,  $a=0.131$ and $b=1.103$
  (see text).}
\end{figure}

\begin{figure}
\includegraphics{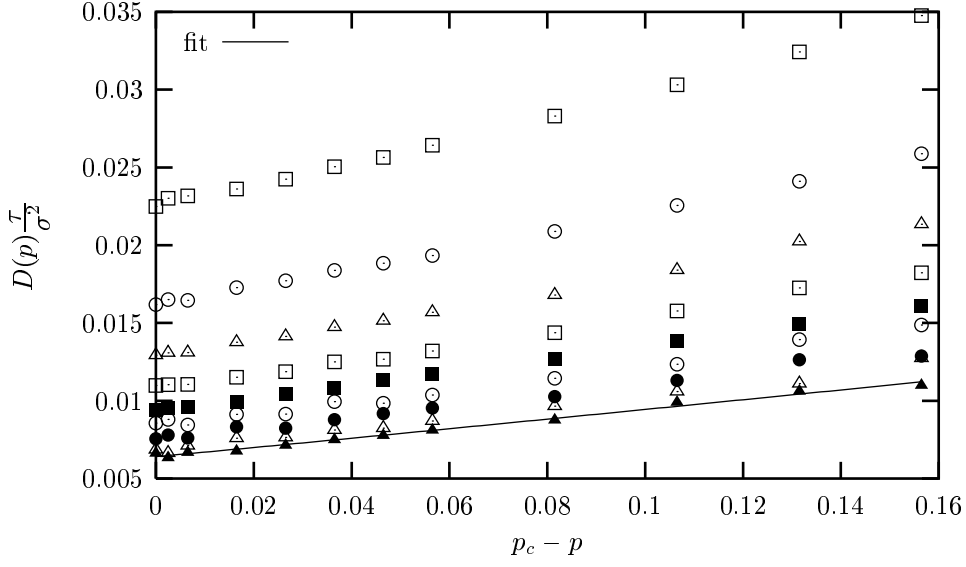}
\caption{\label{fig.Dvsp2} Same as \fig{fig.Dvsp}, but for clusters of
  sizes $m=2\ldots 10$ from top to bottom. The solid line is a fit,
  and here  $D_c=0.0064$, $a=0.0324$ and $b=1.029$.}
\end{figure}

However, even in this regime one would expect some additional $p$
dependence of the diffusion coefficient, a point not addressed in
\cite{deGennes79}.   To this end,  we plot in  \fig{fig.Dvsp} $D(m,p)$
as a function of $p$ for monomers,  dimers and trimers,  and  we see
that the diffusion constants decrease (almost linearly) as a function of
$p$.  Moreover the curves seem to fit nicely to the functional form
$D(p)=a(p_c-p)^{b}+D_c$,  with a value of the exponent $b=1.1$.
In  \fig{fig.Dvsp2}  we have made a similar plot for masses $m=2\ldots
10$, and the trends observed above appear to carry over to larger
masses.   The curves are roughly parallel, and therefore it is not
unlikely that the value of $b$ is independent of $m$, but we are unable to
confirm this from a fit to the data:  the exact value of the
exponent appears to be very sensitive  to  noise in the data.

\begin{figure}
\includegraphics{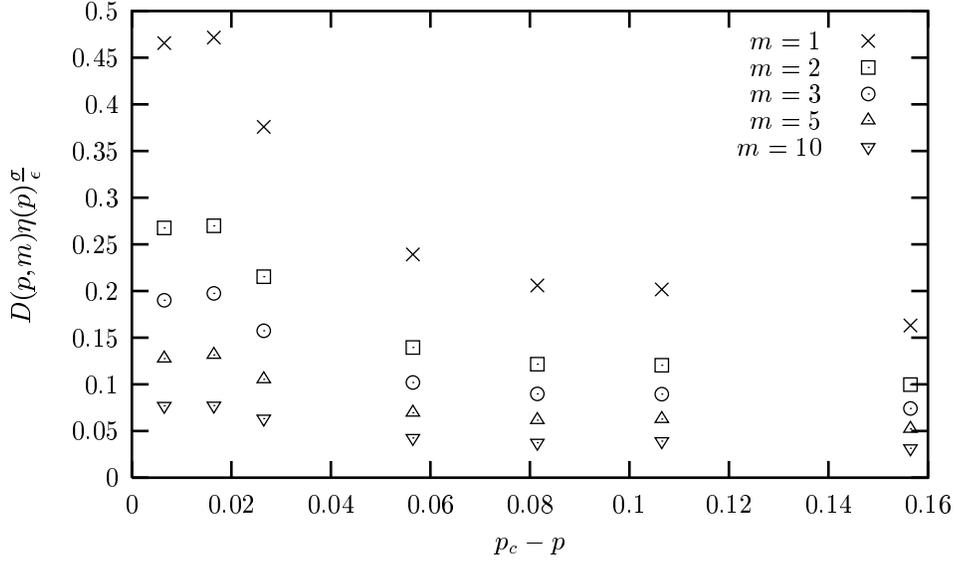}
\caption{\label{fig.Dvsv} Plot of $D(m,p)$ times $\eta$: obviously
  this is only approximately a constant, as predicted by the
 Stokes-Einstein relation, far away from $p_c$.}
\end{figure}

Finally we demonstrate a striking violation of the Stokes-Einstein
relation  when approaching the gelation transition.  The idea that
$D\propto 1/\eta$ is used so widely that one may sometimes forget its
lack of universal validity. In \fig{fig.Dvsv} however it is clear that
$D(m,p)\eta$  increases significantly when $p\to p_c$.
This is consistent with our previous  observations that whereas
$\eta$  diverges  at the gelation point, $D(m,p)$  approaches a
non-vanishing constant  even for large masses $m$.  Further away  from
the gelation point $p\lesssim 0.20$ there does however seem to be an
approximate proportionality between $D(m,p)$  and $\eta$. However, in
Figs.~\ref{fig.Dvsp}--\ref{fig.Dvsp2}  we saw indications that $D(p)\sim
a(p_c-p)^b+D_c$ with $b>1$ whereas $\eta \sim (p_c-p)^{-0.65}$,  and so
this apparent proportionality is at best only approximate.

\section{\label{sec.con}Conclusions}
The main results of this study are the power law behavior of the mass
dependent diffusion coefficient which seems to hold well away from the
critical point, the failure of the Stokes-Einstein relation and
the result $s\approx 0.65$ for the critical exponent of the shear viscosity.
This last result, taken together with other recent results
\cite{vernon01,plischke02} seems  to support the conjecture that the gelation
transition is not classifiable in terms of a single universality class:
exponents in the range $0.3\leq s\leq 0.7$ have been found for models that
seem, on the surface, to be very similar. The experimental situation also
does not provide much evidence for universality: both exponents
near $s=0.7$ \cite{adam81,adam85,durand87} and in the range $1.1\leq s\leq 1.3$
\cite{lusignan95,martin88a,adolf90,martin91,martin88b} have been reported.
We sound a note of caution here: The
determination of exponents through finite size scaling is not very
precise, especially when quantities that are as difficult to calculate
as the shear viscosity form the data set. However, it seems very unlikely
that the errors are large enough that a factor of more than 2 in the
exponent could be explained that way.\par
The mass-dependent diffusion coefficient in this model displays a power
law behavior $D(m,p)\sim m^{-0.69}$ consistent with a scaling argument
of de Gennes \cite{deGennes79}. Reexpressing this in terms of the radius
of gyration of clusters through $m=R_g^{D_f}$ where $D_f=2.5$ is the fractal
dimension of the percolating cluster, the scaling prediction is $D(R_g,p)\sim
R_g^{-(1+s/\nu)}$. This yields an estimate $s=0.65$ for the viscosity exponent,
in good agreement with the direct calculation from the Green-Kubo
formula. Whether this connection between diffusion and viscosity is general
or specific to the present model  and whether there exist a similar
relationship  between diffusion and the elastic shear modulus in the
solid phase remains a subject for further study.
Using a quite different model, del Gado {\it et al.}
\cite{gado00} have studied the self diffusion of crosslinked polymer
clusters on a lattice by bond fluctuation dynamics. They have also used
this scaling ansatz to infer the critical exponent of the shear viscosity
and found $s\approx 1.3$. Their result translates to a mass dependence of
the diffusion constant $D(m,p)\sim m^{-1}$, very different from that of the
present model.\par
Finally, we have shown that as the fluid becomes more viscous there is a
breakdown of the Stokes-Einstein law $D\eta\propto k_BT$ that generally
holds for simple liquids. For relatively small concentrations of crosslinks,
this product varies only very little but for $p\approx p_c$ the divergence of the
viscosity begins to dominate the diffusion constant which seems to saturate
for all cluster sizes studied at $p_c$.

\begin{acknowledgments}
The authors wishes to thank B. J\'oos and D. Vernon for helpful
discussions. Financial support from the Danish National Research
Council grant $21$-$01$-$0335$ and from the NSERC of Canada is gratefully
acknowledged.
\end{acknowledgments}

\bibliography{sfu}

\begin{thebibliography}{24}
\expandafter\ifx\csname natexlab\endcsname\relax\def\natexlab#1{#1}\fi
\expandafter\ifx\csname bibnamefont\endcsname\relax
  \def\bibnamefont#1{#1}\fi
\expandafter\ifx\csname bibfnamefont\endcsname\relax
  \def\bibfnamefont#1{#1}\fi
\expandafter\ifx\csname citenamefont\endcsname\relax
  \def\citenamefont#1{#1}\fi
\expandafter\ifx\csname url\endcsname\relax
  \def\url#1{\texttt{#1}}\fi
\expandafter\ifx\csname urlprefix\endcsname\relax\def\urlprefix{URL }\fi
\providecommand{\bibinfo}[2]{#2}
\providecommand{\eprint}[2][]{\url{#2}}

\bibitem[{\citenamefont{Adam and Lairez}(1996)}]{adam96}
\bibinfo{author}{\bibfnamefont{M.}~\bibnamefont{Adam}} \bibnamefont{and}
  \bibinfo{author}{\bibfnamefont{D.}~\bibnamefont{Lairez}}, in
  \emph{\bibinfo{booktitle}{Physical properties of polymeric gels}}, edited by
  \bibinfo{editor}{\bibfnamefont{J.~P.~C.} \bibnamefont{Addad}}
  (\bibinfo{publisher}{John Wiley \& Sons Ltd}, \bibinfo{year}{1996}),
  p.~\bibinfo{pages}{87}.

\bibitem[{\citenamefont{de~Gennes}(1979)}]{deGennes79}
\bibinfo{author}{\bibfnamefont{P.}~\bibnamefont{de~Gennes}},
  \bibinfo{journal}{J. Phys. Lett. (France)} \textbf{\bibinfo{volume}{40}},
  \bibinfo{pages}{L} (\bibinfo{year}{1979}).

\bibitem[{\citenamefont{Broderix et~al.}(1999)\citenamefont{Broderix, L\"{o}we,
  M\"{u}ller, and Zippelius}}]{broderix99}
\bibinfo{author}{\bibfnamefont{K.}~\bibnamefont{Broderix}},
  \bibinfo{author}{\bibfnamefont{H.}~\bibnamefont{L\"{o}we}},
  \bibinfo{author}{\bibfnamefont{P.}~\bibnamefont{M\"{u}ller}},
  \bibnamefont{and}
  \bibinfo{author}{\bibfnamefont{A.}~\bibnamefont{Zippelius}},
  \bibinfo{journal}{Europhys. Lett} \textbf{\bibinfo{volume}{48}},
  \bibinfo{pages}{421} (\bibinfo{year}{1999}).

\bibitem[{\citenamefont{Jespersen}(2002)}]{sj02}
\bibinfo{author}{\bibfnamefont{S.~N.} \bibnamefont{Jespersen}},
  \bibinfo{journal}{Phys. Rev. E} \textbf{\bibinfo{volume}{66}},
  \bibinfo{pages}{031502} (\bibinfo{year}{2002}).

\bibitem[{\citenamefont{K\"untzel et~al.}(2003)\citenamefont{K\"untzel, L\"owe,
  M\"uller, and Zippelius}}]{kuntzel03}
\bibinfo{author}{\bibfnamefont{M.}~\bibnamefont{K\"untzel}},
  \bibinfo{author}{\bibfnamefont{H.}~\bibnamefont{L\"owe}},
  \bibinfo{author}{\bibfnamefont{P.}~\bibnamefont{M\"uller}}, \bibnamefont{and}
  \bibinfo{author}{\bibfnamefont{A.}~\bibnamefont{Zippelius}},
  \bibinfo{howpublished}{cond-mat/0303578} (\bibinfo{year}{2003}).

\bibitem[{\citenamefont{Smit}(1992)}]{smit92}
\bibinfo{author}{\bibfnamefont{B.}~\bibnamefont{Smit}}, \bibinfo{journal}{J.
  Chem. Phys} \textbf{\bibinfo{volume}{96}}, \bibinfo{pages}{8639}
  (\bibinfo{year}{1992}).

\bibitem[{\citenamefont{Felicity et~al.}(1997)\citenamefont{Felicity, Lodge,
  and Heyes}}]{felicity97}
\bibinfo{author}{\bibfnamefont{J.}~\bibnamefont{Felicity}},
  \bibinfo{author}{\bibfnamefont{M.}~\bibnamefont{Lodge}}, \bibnamefont{and}
  \bibinfo{author}{\bibfnamefont{D.~M.} \bibnamefont{Heyes}},
  \bibinfo{journal}{J. Chem. Soc., Faraday Trans.}  (\bibinfo{year}{1997}).

\bibitem[{\citenamefont{Plischke et~al.}(2002)\citenamefont{Plischke, Jo\'os,
  and Vernon}}]{plischke02}
\bibinfo{author}{\bibfnamefont{M.}~\bibnamefont{Plischke}},
  \bibinfo{author}{\bibfnamefont{B.}~\bibnamefont{Jo\'os}}, \bibnamefont{and}
  \bibinfo{author}{\bibfnamefont{D.}~\bibnamefont{Vernon}}
  (\bibinfo{year}{2002}), \bibinfo{note}{submitted to Phys. Rev. E.}

\bibitem[{\citenamefont{Vernon et~al.}(2001)\citenamefont{Vernon, Plischke, and
  Jo\'os}}]{vernon01}
\bibinfo{author}{\bibfnamefont{D.}~\bibnamefont{Vernon}},
  \bibinfo{author}{\bibfnamefont{M.}~\bibnamefont{Plischke}}, \bibnamefont{and}
  \bibinfo{author}{\bibfnamefont{B.}~\bibnamefont{Jo\'os}},
  \bibinfo{journal}{Phys. Rev. E} \textbf{\bibinfo{volume}{64}},
  \bibinfo{pages}{031505} (\bibinfo{year}{2001}).

\bibitem[{\citenamefont{Ziff and Newman}(2002)}]{ziff02}
\bibinfo{author}{\bibfnamefont{R.~M.} \bibnamefont{Ziff}} \bibnamefont{and}
  \bibinfo{author}{\bibfnamefont{M.~E.~J.} \bibnamefont{Newman}},
  \bibinfo{journal}{Phys. Rev. E} \textbf{\bibinfo{volume}{66}},
  \bibinfo{pages}{016129} (\bibinfo{year}{2002}).

\bibitem[{\citenamefont{Newman and Ziff}(2001)}]{newman01}
\bibinfo{author}{\bibfnamefont{M.~E.~J.} \bibnamefont{Newman}}
  \bibnamefont{and} \bibinfo{author}{\bibfnamefont{R.~M.} \bibnamefont{Ziff}},
  \bibinfo{journal}{Phys. Rev. E} \textbf{\bibinfo{volume}{64}},
  \bibinfo{pages}{016706} (\bibinfo{year}{2001}).

\bibitem[{\citenamefont{Stauffer and Aharony}(1992)}]{stauffer92}
\bibinfo{author}{\bibfnamefont{D.}~\bibnamefont{Stauffer}} \bibnamefont{and}
  \bibinfo{author}{\bibfnamefont{A.}~\bibnamefont{Aharony}},
  \emph{\bibinfo{title}{Introduction to percolation theory}}
  (\bibinfo{publisher}{Taylor \& Francis}, \bibinfo{address}{London},
  \bibinfo{year}{1992}), \bibinfo{edition}{2nd} ed.

\bibitem[{\citenamefont{Allen and Tildesley}(1992)}]{Allen}
\bibinfo{author}{\bibfnamefont{M.~P.} \bibnamefont{Allen}} \bibnamefont{and}
  \bibinfo{author}{\bibfnamefont{D.}~\bibnamefont{Tildesley}},
  \emph{\bibinfo{title}{Computer Simulation of Liquids}}
  (\bibinfo{publisher}{Oxford Science Publications}, \bibinfo{address}{Oxford},
  \bibinfo{year}{1992}).

\bibitem[{\citenamefont{Hansen and McDonald}(1976)}]{Hansen}
\bibinfo{author}{\bibfnamefont{J.~P.} \bibnamefont{Hansen}} \bibnamefont{and}
  \bibinfo{author}{\bibfnamefont{I.~R.} \bibnamefont{McDonald}},
  \emph{\bibinfo{title}{Theory of Simple Liquids}}
  (\bibinfo{publisher}{Academic Press}, \bibinfo{address}{London},
  \bibinfo{year}{1976}).

\bibitem[{\citenamefont{Broderix et~al.}(2001)\citenamefont{Broderix,
  Aspelmeier, Hartmann, and Zippelius}}]{broderix01}
\bibinfo{author}{\bibfnamefont{K.}~\bibnamefont{Broderix}},
  \bibinfo{author}{\bibfnamefont{T.}~\bibnamefont{Aspelmeier}},
  \bibinfo{author}{\bibfnamefont{A.~K.} \bibnamefont{Hartmann}},
  \bibnamefont{and}
  \bibinfo{author}{\bibfnamefont{A.}~\bibnamefont{Zippelius}},
  \bibinfo{journal}{Phys. Rev. E} \textbf{\bibinfo{volume}{64}},
  \bibinfo{pages}{021404} (\bibinfo{year}{2001}).

\bibitem[{\citenamefont{Gado et~al.}(2000)\citenamefont{Gado, de~Arcangelis,
  and Coniglio}}]{gado00}
\bibinfo{author}{\bibfnamefont{E.~D.} \bibnamefont{Gado}},
  \bibinfo{author}{\bibfnamefont{L.}~\bibnamefont{de~Arcangelis}},
  \bibnamefont{and} \bibinfo{author}{\bibfnamefont{A.}~\bibnamefont{Coniglio}},
  \bibinfo{journal}{E. Phys. J. E} \textbf{\bibinfo{volume}{2}},
  \bibinfo{pages}{359} (\bibinfo{year}{2000}).

\bibitem[{\citenamefont{Adam et~al.}(1981)\citenamefont{Adam, Delsanti, Durand,
  Hild, and Munch}}]{adam81}
\bibinfo{author}{\bibfnamefont{M.}~\bibnamefont{Adam}},
  \bibinfo{author}{\bibfnamefont{M.}~\bibnamefont{Delsanti}},
  \bibinfo{author}{\bibfnamefont{D.}~\bibnamefont{Durand}},
  \bibinfo{author}{\bibfnamefont{G.}~\bibnamefont{Hild}}, \bibnamefont{and}
  \bibinfo{author}{\bibfnamefont{J.}~\bibnamefont{Munch}},
  \bibinfo{journal}{Pure Appl. Chem} \textbf{\bibinfo{volume}{53}},
  \bibinfo{pages}{1489} (\bibinfo{year}{1981}).

\bibitem[{\citenamefont{Adam et~al.}(1985)\citenamefont{Adam, Delsanti, and
  Durand}}]{adam85}
\bibinfo{author}{\bibfnamefont{M.}~\bibnamefont{Adam}},
  \bibinfo{author}{\bibfnamefont{M.}~\bibnamefont{Delsanti}}, \bibnamefont{and}
  \bibinfo{author}{\bibfnamefont{D.}~\bibnamefont{Durand}},
  \bibinfo{journal}{Macromolecules} \textbf{\bibinfo{volume}{18}},
  \bibinfo{pages}{2285} (\bibinfo{year}{1985}).

\bibitem[{\citenamefont{Durand et~al.}(1987)\citenamefont{Durand, Delsanti, and
  Luck}}]{durand87}
\bibinfo{author}{\bibfnamefont{D.}~\bibnamefont{Durand}},
  \bibinfo{author}{\bibfnamefont{M.}~\bibnamefont{Delsanti}}, \bibnamefont{and}
  \bibinfo{author}{\bibfnamefont{J.~M.} \bibnamefont{Luck}},
  \bibinfo{journal}{Europhys. Lett} \textbf{\bibinfo{volume}{3}},
  \bibinfo{pages}{297} (\bibinfo{year}{1987}).

\bibitem[{\citenamefont{Lusignan et~al.}(1995)\citenamefont{Lusignan, Mourey,
  Wilson, and Colby}}]{lusignan95}
\bibinfo{author}{\bibfnamefont{C.}~\bibnamefont{Lusignan}},
  \bibinfo{author}{\bibfnamefont{T.}~\bibnamefont{Mourey}},
  \bibinfo{author}{\bibfnamefont{J.~C.} \bibnamefont{Wilson}},
  \bibnamefont{and} \bibinfo{author}{\bibfnamefont{R.~H.} \bibnamefont{Colby}},
  \bibinfo{journal}{Phys. Rev. E} \textbf{\bibinfo{volume}{52}},
  \bibinfo{pages}{6271} (\bibinfo{year}{1995}).

\bibitem[{\citenamefont{Martin and Wilcoxon}(1988)}]{martin88a}
\bibinfo{author}{\bibfnamefont{J.}~\bibnamefont{Martin}} \bibnamefont{and}
  \bibinfo{author}{\bibfnamefont{J.}~\bibnamefont{Wilcoxon}},
  \bibinfo{journal}{Phys. Rev. Lett.} \textbf{\bibinfo{volume}{61}},
  \bibinfo{pages}{373} (\bibinfo{year}{1988}).

\bibitem[{\citenamefont{Adolf and Martin}(1990)}]{adolf90}
\bibinfo{author}{\bibfnamefont{D.}~\bibnamefont{Adolf}} \bibnamefont{and}
  \bibinfo{author}{\bibfnamefont{J.}~\bibnamefont{Martin}},
  \bibinfo{journal}{Macromolecules} \textbf{\bibinfo{volume}{23}},
  \bibinfo{pages}{3700} (\bibinfo{year}{1990}).

\bibitem[{\citenamefont{Martin et~al.}(1991)\citenamefont{Martin, Wilcoxon, and
  Odinek}}]{martin91}
\bibinfo{author}{\bibfnamefont{J.}~\bibnamefont{Martin}},
  \bibinfo{author}{\bibfnamefont{J.}~\bibnamefont{Wilcoxon}}, \bibnamefont{and}
  \bibinfo{author}{\bibfnamefont{J.}~\bibnamefont{Odinek}},
  \bibinfo{journal}{Phys. Rev. A} \textbf{\bibinfo{volume}{43}},
  \bibinfo{pages}{858} (\bibinfo{year}{1991}).

\bibitem[{\citenamefont{Martin et~al.}(1988)\citenamefont{Martin, Wilcoxon, and
  Adolf}}]{martin88b}
\bibinfo{author}{\bibfnamefont{J.}~\bibnamefont{Martin}},
  \bibinfo{author}{\bibfnamefont{J.}~\bibnamefont{Wilcoxon}}, \bibnamefont{and}
  \bibinfo{author}{\bibfnamefont{D.}~\bibnamefont{Adolf}},
  \bibinfo{journal}{Phys. Rev. Lett.} \textbf{\bibinfo{volume}{61}},
  \bibinfo{pages}{2620} (\bibinfo{year}{1988}).

\end{thebibliography}
\end{document}